\documentclass[twocolumn,showpacs,amsmath,amssymb,superscriptaddress,nofootinbib]{revtex4}

\usepackage{graphicx}
\usepackage{dcolumn}
\usepackage{bm}

\def\totacctxt{4.0\times 10^{-6}}
\def\totacc{4.01\times 10^{-6}}
\def\sescomb{1.07\times 10^{-6}}
\def\sysuncp{4.9}
\def\sessys{(1.07\pm 0.05(syst.)) \times 10^{-6}}

\def\brses{(2.51 \pm 0.74(stat.) \pm 0.12(syst.))\times 10^{-5}}

\def\brfinal{(1.58 \pm  0.46(stat.) \pm 0.08(syst.) ) \times 10^{-5}}

\def\brpdg{(1.10 \pm 0.32(stat.) \pm 0.05(syst.))\times 10^{-5}}

\begin{document}

\title{
Measurement of the $K^{+}\to\pi^{0}\mu^{+}\nu_{\mu}\gamma$ Branching Ratio
}

\author{S.~Adler} \affiliation{Brookhaven National Laboratory, Upton,
New York 11973} 

\author{A.O.~Bazarko} \altaffiliation{Present address: 
 Schlumberger, Princeton Jct., NJ 08550.}
\affiliation{Joseph Henry Laboratories, Princeton University, 
Princeton, New Jersey 08544}

\author{P.C.~Bergbusch} \affiliation{Department of Physics and
Astronomy, University of British Columbia, Vancouver, British
Columbia, Canada, V6T 1Z1} 

\author{E.W.~Blackmore}
\affiliation{TRIUMF, 4004 Wesbrook Mall, Vancouver, British Columbia,
Canada, V6T 2A3} 

\author{D.A.~Bryman} \affiliation{Department of
Physics and Astronomy, University of British Columbia, Vancouver,
British Columbia, Canada, V6T 1Z1} 

\author{S.~Chen} \altaffiliation{Present address: 
Department of Engineering Physics, Tsinghua University,
Beijing 100084, China.}
\affiliation{TRIUMF, 4004 Wesbrook Mall, Vancouver, British Columbia,
Canada, V6T 2A3}

\author{I-H.~Chiang} \affiliation{Brookhaven National
Laboratory, Upton, New York 11973} 

\author{M.V.~Diwan}
\affiliation{Brookhaven National Laboratory, Upton, New York 11973}

\author{J.S.~Frank} \affiliation{Brookhaven National Laboratory,
Upton, New York 11973} 

\author{T.~Fujiwara}
\email{fujiwara@scphys.kyoto-u.ac.jp}
\affiliation{Department of Physics, Kyoto University. Kitashirakawa, Sakyo, Kyoto 606-8502, Japan}

\author{J.S.~Haggerty} \affiliation{Brookhaven National Laboratory, 
Upton, New York 11973} 

\author{J.~Hu} \affiliation{TRIUMF, 4004 Wesbrook Mall, Vancouver, British Columbia,
Canada, V6T 2A3}

\author{T.~Inagaki} \affiliation{High Energy
Accelerator Research Organization (KEK), Oho, Tsukuba, Ibaraki
305-0801, Japan} 

\author{M.M.~Ito}\altaffiliation{Present address:
Thomas Jefferson National Accelerator Facility, 
Newport News, VA 23606.}
\affiliation{Joseph Henry Laboratories,
Princeton University, Princeton, New Jersey 08544}

\author{D.E.~Jaffe}
\affiliation{Brookhaven National Laboratory, Upton, New York 11973}

\author{V.~Jain}\altaffiliation{Present address:
Department of Physics, Indiana University,
Bloomington, IN 47405-7105.}
\affiliation{Brookhaven National Laboratory, Upton, New York 11973}

\author{S.~Kabe} \affiliation{High Energy Accelerator Research
Organization (KEK), Oho, Tsukuba, Ibaraki 305-0801, Japan}

\author{S.H.~Kettell} \affiliation{Brookhaven National Laboratory,
Upton, New York 11973}
 
\author{P.~Kitching} \affiliation{Centre for
Subatomic Research, University of Alberta, Edmonton, Canada, T6G 2N5}

\author{M.~Kobayashi} \affiliation{High Energy Accelerator Research
Organization (KEK), Oho, Tsukuba, Ibaraki 305-0801, Japan}

\author{T.K.~Komatsubara} \affiliation{High Energy Accelerator
Research Organization (KEK), Oho, Tsukuba, Ibaraki 305-0801, Japan}

\author{A.~Konaka} \affiliation{TRIUMF, 4004 Wesbrook Mall, Vancouver,
British Columbia, Canada, V6T 2A3} 

\author{Y.~Kuno}\altaffiliation{Present address: Department of Physics, Osaka
University, Toyonaka, Osaka 560-0043, Japan.} 
\affiliation{High Energy Accelerator Research Organization (KEK), Oho, Tsukuba, Ibaraki
305-0801, Japan} 

\author{M.~Kuriki} \altaffiliation{Present address: Graduate School of
Advanced Sciences of Matter, Hiroshima University, Hiroshima 739-8530, Japan.}
\affiliation{High Energy
Accelerator Research Organization (KEK), Oho, Tsukuba, Ibaraki
305-0801, Japan} 

\author{K.K.~Li} \affiliation{Brookhaven National Laboratory,
Upton, New York 11973}

\author{L.S.~Littenberg} \affiliation{Brookhaven National Laboratory,
Upton, New York 11973} 

\author{J.A.~Macdonald} \altaffiliation{Deceased}\affiliation{TRIUMF,
4004 Wesbrook Mall, Vancouver, British Columbia, Canada, V6T 2A3}

\author{ P.D.~Meyers} \affiliation{Joseph Henry Laboratories,
Princeton University, Princeton, New Jersey 08544}

\author{J.~Mildenberger} \affiliation{TRIUMF, 4004 Wesbrook Mall,
Vancouver, British Columbia, Canada, V6T 2A3} 

\author{M.~Miyajima} \affiliation{Department of Applied Physics, 
Fukui University, 3-9-1 Bunkyo, Fukui, Fukui 910-8507, Japan}

\author{N.~Muramatsu}
\affiliation{Research Center for Nuclear Physics, Osaka University,
10-1 Mihogaoka, Ibaraki, Osaka 567-0047, Japan} 

\author{T.~Nakano}
\affiliation{Research Center for Nuclear Physics, Osaka University,
10-1 Mihogaoka, Ibaraki, Osaka 567-0047, Japan} 

\author{C.~Ng} \altaffiliation{Also at Physics Department, State University of New
York at Stony Brook, Stony Brook, NY 11794-3800.}
\affiliation{Brookhaven National Laboratory, Upton, New York 11973}

\author{S.~Ng} \affiliation{Centre for
Subatomic Research, University of Alberta, Edmonton, Canada, T6G 2N5}

\author{T.~Nomura} \altaffiliation{Present address: High Energy Accelerator
Research Organization (KEK).}
\affiliation{Department of Physics, Kyoto University. Kitashirakawa, Sakyo, Kyoto 606-8502, Japan}

\author{T.~Numao} \affiliation{TRIUMF, 4004 Wesbrook Mall, Vancouver,
British Columbia, Canada, V6T 2A3} 

\author{J.-M.~Poutissou}
\affiliation{TRIUMF, 4004 Wesbrook Mall, Vancouver, British Columbia,
Canada, V6T 2A3} 

\author{R.~Poutissou} \affiliation{TRIUMF, 4004
Wesbrook Mall, Vancouver, British Columbia, Canada, V6T 2A3}

\author{G.~Redlinger} \altaffiliation{Present address: Brookhaven National Laboratory.}
\affiliation{TRIUMF, 4004 Wesbrook Mall, Vancouver, British Columbia,
Canada, V6T 2A3} 

\author{T.~Sato} \affiliation{High Energy Accelerator
Research Organization (KEK), Oho, Tsukuba, Ibaraki 305-0801, Japan}

\author{K.~Shimada} \affiliation{Department of Applied Physics, 
Fukui University, 3-9-1 Bunkyo, Fukui, Fukui 910-8507, Japan}

\author{T.~Shimoyama} \affiliation{Department of Applied Physics, 
Fukui University, 3-9-1 Bunkyo, Fukui, Fukui 910-8507, Japan}

\author{T.~Shinkawa} \altaffiliation{Present address: Department of Applied
Physics, National Defense Academy, Yokosuka, Kanagawa 239-8686, Japan.}
\affiliation{High Energy Accelerator Research Organization (KEK), Oho,
Tsukuba, Ibaraki 305-0801, Japan} 

\author{F.C.~Shoemaker}\altaffiliation{Deceased}
\affiliation{Joseph Henry Laboratories, Princeton University,
Princeton, New Jersey 08544} 

\author{J.R.~Stone} \affiliation{Joseph
Henry Laboratories, Princeton University, Princeton, New Jersey 08544}

\author{R.C.~Strand} \affiliation{Brookhaven National Laboratory,
Upton, New York 11973} 

\author{S.~Sugimoto} \affiliation{High Energy
Accelerator Research Organization (KEK), Oho, Tsukuba, Ibaraki
305-0801, Japan} 

\author{Y.~Tamagawa} \affiliation{Department of Applied Physics, 
Fukui University, 3-9-1 Bunkyo, Fukui, Fukui 910-8507, Japan}

\author{T.~Tsunemi} \altaffiliation{Present address: Department of Physics, 
Kyoto University.}
\affiliation{High Energy
Accelerator Research Organization (KEK), Oho, Tsukuba, Ibaraki
305-0801, Japan} 

\author{C.~Witzig} \affiliation{Brookhaven National
Laboratory, Upton, New York 11973} 

\author{Y.~Yoshimura}
\affiliation{High Energy Accelerator Research Organization (KEK), Oho,
Tsukuba, Ibaraki 305-0801, Japan}

\collaboration{E787 Collaboration}

\date{April 9, 2010}

\begin{abstract}
   A measurement of the decay
   $K^{+}\to\pi^{0}\mu^{+}\nu_{\mu}\gamma$
   has been performed 
  with the E787 detector 
  at Brookhaven National Laboratory.
  Forty events were observed in the signal region 
  with the background expectation of $(16.5\pm 2.7)$ events.
 The branching ratio 
 was measured to be
 $\brfinal$
  in the kinematic region
 $E_{\gamma}>30$ MeV and  $\theta_{\mu\gamma}>20^{\circ}$,
 where $E_{\gamma}$ is the energy of the emitted photon and
 $\theta_{\mu\gamma}$ is the angle between the muon and the photon
in the $K^+$ rest frame.
The results were consistent with theoretical predictions. 
 \end{abstract}


\maketitle

\section{Introduction}
\label{sec-introduction}

    A measurement of the decay 
    $K^{+}\to\pi^{0}\mu^{+}\nu_{\mu}\gamma$
    ($K_{\mu3\gamma}$)
    has been performed 
    using the E787 apparatus \cite{e787web}
    at the Alternating Gradient Synchrotron (AGS)
    of Brookhaven National Laboratory.
    This decay is due to radiative effects 
    in the semi-leptonic transition 
    of $K^{+}\to\pi^{0}\mu^{+}\nu_{\mu}$ decay ($K_{\mu3}$). 
    The decay $K_{\mu3\gamma}$  can proceed via 
    inner bremsstrahlung (IB) in which a photon is emitted 
    from the charged particle in the initial or final state,
    and structure-dependent radiative decay (SD)
    which involves the emission of a photon 
     from the intermediate states 
     in the hadronic transition from $K^+$ to $\pi^{0}$.
    For hadron decays
    in the energy region below 1 GeV,
    the effective-field approach based on chiral symmetry,  
    chiral perturbation theory (ChPT) \cite{ChPT},
    is applicable \cite{Holstein90}. 
    At $O(p^4)$ in ChPT, 
    a partial branching ratio
    $BR(K_{\mu3\gamma},\ E_{\gamma} > 30$ MeV and $\theta_{\mu\gamma}>20^{\circ})$,
    where 
    $E_{\gamma}$ is the energy of the emitted photon
    and
    $\theta_{\mu\gamma}$ is the angle between the muon and the photon
    in the $K^+$ rest frame,
    is predicted to be $2.0\times 10^{-5}$ \cite{BEG-DAFNE, BEG93}.
    The relative size of SD contribution is around 8 \%.
    
    The first search for $K_{\mu3\gamma}$, which was performed  in 1973 \cite{Ljung73}
    with $K^+$ decays at rest in  a heavy-liquid bubble chamber,
    obtained 
    $BR(K_{\mu3\gamma},\ E_{\gamma} > 30$ MeV$)<6.1\times 10^{-5}$
    at the 90\% confidence level (C.L.).
    The  $K^{-}\to\pi^{0}\mu^{-}\nu_{\mu}\gamma$  decay in flight 
    was studied
    by the ISTR+ spectrometer; 
    the ratio $BR(K_{\mu3\gamma})/BR(K_{\mu 3})$
    for the region $30 < E_{\gamma} < 60$ MeV
    was found to be $(4.48 \pm 0.68(stat.)\pm 0.99(syst.))\times 10^{-4}$ \cite{ISTR}.
     Based on this value, 
     the Particle Data Group cited 
     $BR(K_{\mu3\gamma},\ 30 < E_{\gamma} < 60$ MeV$)$ as 
     $(1.5\pm 0.4) \times 10^{-5}$   \cite{PDG08}.
    The  $K^{+}\to\pi^{0}\mu^{+}\nu_{\mu}\gamma$  decay at rest was studied
    with the toroidal spectrometer 
    of the E246 and E470 experiments at the KEK 12-GeV proton synchrotron;
    the $BR(K_{\mu3\gamma},\ E_{\gamma} > 30$ MeV and $\theta_{\mu\gamma}>20^{\circ})$
    was found to be
    $(2.4\pm 0.5(stat.) \pm 0.6(syst.)) \times 10^{-5}$ \cite{Shimizu06}.

\begin{figure*}
 \includegraphics[scale=0.50]{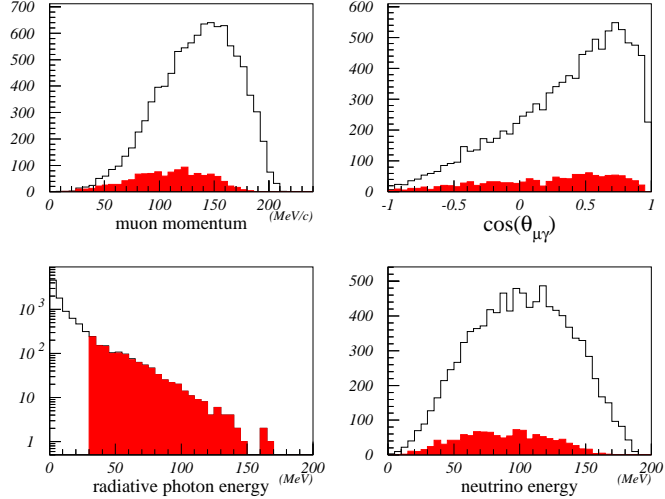}
 \caption{Spectra of $K^+\to\pi^0\mu^+\nu_{\mu}\gamma$ observables 
 based on ChPT at $O(p^4)$:
 muon momentum (top left), cosine of $\theta_{\mu\gamma}$ (top right),   
 $E_{\gamma}$ (bottom left) and neutrino energy (bottom right).
 The unhatched and hatched histograms represent
 the distributions before and after imposing the conditions 
 $E_{\gamma} > 30$ MeV and $\theta_{\mu\gamma}>20^{\circ}$.}
 \label{fig_theory}
 \end{figure*}

   In this paper, we present results from the E787 experiment at the AGS. 
   We describe the analysis we have performed
   to observe the  $K^{+}\to\pi^{0}\mu^{+}\nu_{\mu}\gamma$  decay
   and the measurement of 
   the partial branching ratio 
   $BR(K_{\mu3\gamma},\ E_{\gamma} > 30$ MeV and $\theta_{\mu\gamma}>20^{\circ})$.
   We compare our results to the predictions of ChPT at $O(p^4)$.
   The $BR(K_{\mu3\gamma},\ 30 < E_{\gamma} < 60$ MeV$)$ 
   was also determined and can be compared with the previous result.
   The spectra of $K_{\mu3\gamma}$ observables 
   based on ChPT at $O(p^4)$ \cite{BEG-DAFNE, BEG93}
   are shown in Fig.~\ref{fig_theory}
   and are consistent with the spectra given in \cite{Tviol2}.

\section{Detector}
\label{sec-detector}

\begin{figure*}
 \includegraphics[scale=0.5]{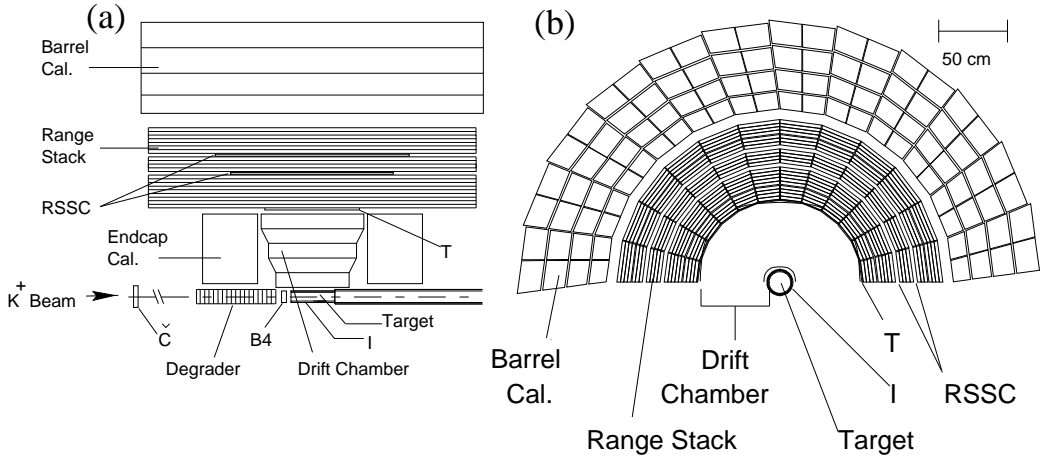}
 \caption{Schematic side(a) and end(b) views showing the upper half 
 of the E787 detector. \v{C}: \v{C}erenkov counter; B4: beam counter;
 I and T; trigger scintillation counters (I-counters and T-counters).
 Range-stack straw-tube tracking chambers (RSSC) were unused in this measurement.}
 \label{fig_detector}
 \end{figure*}

  E787 was a rare kaon-decay experiment studying 
  $K^+\to\pi^+\nu\bar{\nu}$ \cite{E787-959697,E787-pnn2,E787-98}
  and related decays \cite{E787-kp2g,E787-bypro}
  using $K^+$ decays at rest.
  The data for the $K_{\mu3\gamma}$ measurement
  were acquired during the 1998 run \cite{E787-98} of the AGS,
  using kaons of 710 MeV/$c$ incident on the E787 apparatus 
  at a rate of about 4 MHz.
  Two stages of electrostatic particle separation
  in the beam line \cite{LESB3}
  reduced the pion contamination at the E787 apparatus to 25\%. 
  The incident kaons were detected and identified by 
  a \v{C}erenkov counter, multi-wire proportional chambers and a beam counter.
  After being slowed by a BeO degrader, 
  27\% of the kaons  came to rest in an active stopping target
  located at the center of the detector system \cite{e787det}
  in a 1.0 Tesla solenoidal magnet
  (Fig.~\ref{fig_detector}).    

   The 12-cm diameter target, 
   which consisted of 0.5-cm square plastic scintillating fibers,
   provided initial tracking of the stopping kaon
   as well as  the charged decay products.
   They passed through 
   a 0.64-cm thick layer of plastic scintillation counters
   surrounding the target (I-counters).
   A  $3\times 10^{-2}$ radiation length ($X_0$) thick
   cylindrical drift-chamber \cite{UTC} 
   provided tracking information for momentum determination.   
   The charged particles lost energy in an array of plastic scintillation counters 
   called the Range Stack (RS), which 
   provided a measurement of kinetic energy and range 
   (equivalent cm of plastic scintillator) 
   of the charged track.
   The RS was segmented into 24 azimuthal sectors and
   21 radial layers, totaling one $X_0$.
   The RS counters in the first layer (T-counters), 
   which were 0.635-cm thick and 52-cm long,
   defined the solid angle acceptance
   for the charged track in the RS. 
   The subsequent RS counters in the second layer (L2-counters) and beyond
   were 1.905-cm thick and 182-cm long.
   The RS counter in the sector and layer where the track came to rest
   was called 
   the ``stopping counter''.   
   The RS counters were read out by phototubes 
   attached to the upstream and downstream ends; 
   the output pulse shapes were recorded by
   500-MHz sampling transient digitizers (TDs) \cite{TD},
   which provided precise time and energy information
   for the hits in each counter.

   Photons from kaon decays at rest in the target
   were detected by a hermetic calorimeter system
   surrounding the central region. 
   The cylindrical barrel (BL) calorimeter, 
   which was located immediately outside the RS 
   and covered two thirds of the solid angle, 
   was used to reconstruct and measure 
   the three photons  from the $K_{\mu3\gamma}$ decay
   (i.e. two photons from the $\pi^0$ and the radiative photon).
   The BL calorimeter consisted of
   alternating layers of lead (0.1-cm thick) and plastic scintillator (0.5-cm thick) sheets,
   and about 29\% of the electromagnetic shower energy was deposited in the scintillators.
   The BL was segmented azimuthally into 48 sectors and,
   in each sector, four radial groups of 16, 18, 20 and 21 lead-scintillator layers,
   respectively with increasing radius, formed BL modules 
   totaling 14.3 $X_0$.
   The modules were 190-cm in length along the beam axis.
   They were read out by phototubes attached to the upstream and downstream ends, and
   the outputs were recorded by time-to-digital converters (TDCs) and 
   analogue-to-digital  converters (ADCs).
   The two endcap calorimeters \cite{CsI}, 
   additional ``collar" calorimeters 
   for filling minor openings 
   along the beam direction, 
   and any active parts of the detector not hit by the charged track 
   were used for detecting extra particles.

\section{Trigger}
\label{sec-trigger}

    The signature of the $K_{\mu3\gamma}$ decay was
    a kaon decay at rest 
    with a charged track in the RS and three electromagnetic showers 
    in the BL calorimeter.
    The kinetic energy of most of the muons from $K_{\mu3\gamma}$
    after requiring  $E_{\gamma} > 30$ MeV and $\theta_{\mu\gamma}>20^{\circ}$
    (Fig.~\ref{fig_theory})
    is less than 85 MeV; 
    thus, we used the RS counters in the first six layers for the muon measurement. 
    The RS counters in the 7th layer and beyond were used 
    to detect extra particles
    or reject photons whose showers started before reaching the BL.
 
    In the trigger for the $K_{\mu3\gamma}$ decay, 
    the ``3$\gamma$ trigger'', 
    a kaon was identified by a coincidence of hits from the \v{C}erenkov counter,
    beam counter and target.
    To guarantee that the kaon actually decayed at rest, 
    the timing of  the outgoing particle (via the I-counters) was required 
    to be at least 1.5 nsec later
    than the timing of the incoming kaon (via the \v{C}erenkov counter). 
    A positively charged track was required to have a coincidence of the hits from the I-counters, 
    a T-counter and an L2-counter in the same RS sector (T$\cdot$2 sector), 
    and from the counters in the 3rd or 4th RS layer 
    in the T$\cdot$2 sector
    or in either of the next two clockwise sectors.
    The tracks reaching the 7th or 8th layer were rejected.
    With such trigger conditions, 
    only tracks for which the RS stopping counter
    was located in the 3rd to the 6th layer,
    corresponding to the muons with momentum 100-160 MeV/$c$,
    were accepted.

    In order to count the number of showers in the BL calorimeter online,
    the analog sum of the outputs from 8 modules in two adjacent sectors, 
    separately for the upstream and downstream ends, were accepted 
    with a threshold 
    corresponding to about 5 MeV of visible energy per end.
    The discriminator outputs were OR-ed for either end, 
    and then fed into a logic unit.
    The number of showers 
    was required to be equal to three or larger.
    An event was rejected online
    if the visible energy in the endcap calorimeters was more than 20 MeV, 
    the energy in the RS sectors outside the region of the charged track 
    was more than 10 MeV,
    or the energy in the RS counters from the 9th to 21st layers in any sector
    was larger than the level for minimum ionizing particles.

    The 3$\gamma$ trigger was prescaled by five for taking data simultaneously 
    with the trigger for $K^+\to\pi^+\nu\bar{\nu}$.
    Approximately 6 events per slow-extracted beam (2.2-sec duration) from the  AGS in every 4.2 sec
    were recorded. 
    A total exposure of kaons entering the target available for the $K_{\mu3\gamma}$ measurement
    was $3.5\times10^{11}$.
   A total of $9.4\times10^{6}$ triggered events were collected.

   Monte Carlo simulation was performed 
   to generate samples of $K_{\mu3\gamma}$ 
   as well as other kaon decays at rest
   in the E787 detector system. The simulation package developed for E787 
   was comprised of 
   the electromagnetic-shower simulation routines from EGS4~\cite{EGS4}, 
   the routines to model other physics processes such as particle decays
   and charged-particle interactions in the detector elements,
   and the routines to simulate the trigger conditions. 
   The events in the Monte Carlo samples can be analyzed with the same analysis codes
   for real data, 
   except that the information on the pulse shapes in the digitizers 
   and the information on the outputs from most of the beam instrumentation
   were not stored. 
   The  Monte Carlo simulation of E787 is described in \cite{pnn1full}.

\section{Event reconstruction}
\label{sec-reconstruction}

\subsection{Basic reconstruction}
\label{subsec-basic}

  The charged track  was reconstructed and measured 
  with a standard algorithm used in the $K^+\to\pi^+\nu\bar{\nu}$ analysis
  of E787 described in \cite{pnn1full}, 
  except that the charged particle was assumed to be a muon and 
  all the energy deposited in the RS stopping counter 
  was attributed to the muon track coming to rest~\footnote{
       In the $\pi^+$ track reconstruction, the presence of the additional 4 MeV deposit
       in the RS stopping counter
       due to  the $\pi^+\to\mu^+\nu$ decay at rest 
       was assumed and was subtracted from the observed energy.}.
  The initial momentum was determined
  by correcting the momentum measured in the drift chamber
  for the energy loss suffered by the muon
  based on the observed track length in the target.
  The kinetic energy was determined by adding up the energy deposits of 
  the muon track in the scintillators of the target and the RS,
  taking account of the energy loss in inactive materials
  such as wrapping and chamber materials.
  The range was calculated from the track length in the target and in the RS, 
  and was used for muon identification 
  in the $K_{\mu 3\gamma}$ analysis.
  Events were discarded
  if the momentum of the charged particle was larger than 190 MeV/$c$.
  Many of the triggered events were due to 
  the $K^+\to\pi^+\pi^0$ decay for which the $\pi^+$ had interactions 
  within the RS counters before coming to rest; 
  these events were easily rejected
  because the $\pi^+$ momentum (205 MeV/$c$) was measured
  with the drift chamber.

  The isolated showers
  in the BL calorimeter
  were reconstructed
  and the timing, energy and direction
  of the photons were measured 
  with the same algorithm used in the analysis of
  E787 and E949 
  (the successor experiment to E787) \cite{pnn1full,E949pnn1,E9491g,E949pv,E949pnn2}
  for kaon decays with photons 
  in the final states \cite{E787-kp2g,E787-bypro,E9491g}. 
  The muon track in the RS defined the event time reference, and 
  adjacent BL modules 
  whose timing was within $\pm 6$ nsec of the track time and
  whose visible energy was more than 0.2 MeV
  were grouped into a ``cluster".
  In each module,
  the energy was measured as the geometrical mean 
  of the visible energies
  divided by the calibrated visible fraction ($0.292$)
  in the upstream and downstream ends. 
  The hit position in the end view of the detector was determined
  by the segmentation of the modules;  
  the hit position along the beam axis ($z$) was measured
  from the end-to-end time and energy differences. 
  The location of the shower in the BL was obtained by taking an energy-weighted average 
  of the hit positions in the same cluster
  and, in conjunction with the kaon decay vertex position in the target, 
  the azimuthal angle 
  and the polar angle with respect to the beam axis
  were determined for each photon.
  In the $K_{\mu 3\gamma}$ analysis,
  if two showers were separated by less than 55 cm and 
  the energy of one of them was less than 25 MeV, 
  they were combined to a single shower
  in order to avoid misidentifying a part of an electromagnetic shower to be a separate one.
  Events were discarded
  unless three or four showers were observed  in the BL calorimeter.

\subsection{Primary selection}
\label{subsec-primary}

  Selection criteria (``cuts") were imposed 
  to make sure that the trigger conditions were satisfied 
  and the kaon in the initial state and the muon and photons in the final state
  were reconstructed
  within the fiducial volume of the detector system.
  Furthermore, 
  cuts on the timing and energy of the hits recorded 
  in the \v{C}erenkov counter, proportional chambers, 
  beam counter 
  and the target, 
  including an  offline delayed-coincidence cut which required $>2$ nsec
  between the muon  time and the kaon time measured in the target,
  were imposed to remove the events triggered by kaon decays in flight or
  by multiple beam particles into the detector. 
  From the triggered events, 
  $1.1\times 10^{6}$ events survived and were used in the subsequent analysis.

\section{Background sources}
\label{sec-sources}

Kaon decays with a single charged track in the region 100-160 MeV/$c$
and at least one $\pi^0$ in the final state, i.e.,  
$K_{\mu 3}$,  
$K^+\to\pi^+\pi^0\gamma$ ($K_{\pi 2\gamma}$), 
$K^{+}\to \pi^{0}e^{+}\nu_{e}$ ($K_{e 3}$), 
$K^{+}\to \pi^{0}e^{+}\nu_{e}\gamma$ ($K_{e 3 \gamma}$)
and $K^+\to\pi^+\pi^0\pi^0$ ($K_{\pi 3}$), 
are potential background sources. 
They are classified as follows: 
 \begin{description}
  \item[``$K_{\mu 3}$-1":]
  $K_{\mu3}$ decays associated with an extra cluster in the BL
            due to accidental hits.
  \item[``$K_{\mu 3}$-2":]
  $K_{\mu3}$ decays associated with an extra cluster in the BL
  when  the showers due to the two photons from $\pi^0$ were reconstructed
  as three showers.
\item[``$K_{\pi 2\gamma}$":] 
  $K_{\pi 2\gamma}$ decay events in which the $\pi^+$ was misidentified as a muon
  or decayed in flight before it came to rest in the RS.
 \item[``$K_{e 3(\gamma)}$":]
   $K_{e 3 \gamma}$ decays
   for which the $e^{+}$ was misidentified as a muon, 
   or
   $K_{e 3}$ decays with the $e^+$ misidentification
   and an extra cluster in the BL
   or a photon due to bremsstrahlung.
 \item[``$K_{\pi 3}$":]  $K_{\pi 3}$ decays
   in which the $\pi^+$ was misidentified as a muon
   or decayed in flight before it came to rest in the RS,
   and at the same time one of the four photons from two $\pi^0$'s was undetected.
  \end{description}
  
  Background
  due to multiple beam particles scattering into the detector
   simultaneously was found to be negligible. 
   
In the subsequent two sections, 
the cuts to improve the signal-to-background
and measure the background are described.
The cuts were developed 
with the sample of real data prescaled by three 
and the Monte Carlo samples.
The studies of these samples confirmed that 
the expected number of candidate events in the signal region 
was larger than the background expectation. 
The cuts were then imposed to the entire dataset.


\section{Event selection}
\label{sec-offline}

\subsection{Selection of the muon and three photons}
\label{subsec-selection}

For muon identification, 
a cut based on the measured range compared to that expected
given the measured muon momentum 
was imposed. 
The measured range was also compared to  the measured kinetic energy. 
Furthermore, to reject $e^+$ tracks in the RS, 
a set of cuts that checked the consistency between the measured energy
and range in each of the RS counters, the $dE/dx$ cuts, 
were employed.

For the $K_{\mu 3\gamma}$ decay,
events were discarded unless 
exactly three showers with $> 5$ MeV were observed in the BL calorimeter
within $\pm 2$ nsec of the muon track time in the RS.
The photon energy
should also be sufficiently larger than the online energy threshold;
since the threshold had been applied 
to the analog-sum signals  separately for the upstream and downstream ends
(Section~\ref{sec-trigger}), 
the threshold level 
depended on the $z$ position of the shower in the BL.
Thus the photon energy was required to be 
greater than 23 (66) MeV 
for the showers located at the edge (center) of the BL.

In order to remove the events with the photon
due to bremsstrahlung of the $e^+$ track
from the $K_{e 3}$ or $K_{e 3 \gamma}$ decay
within the RS,
 we calculated the angle between 
 the direction of the hit position of the charged track at the RS T-counter 
  from the center of the detector system
 and 
 the direction of each of the photons 
  from the kaon decay vertex position.
The minimum of the angles
should be  larger than $26^{\circ}$
(``$e^+$ bremsstrahlung" cut).

An extra photon could disappear
through inefficiency due to very narrow gaps between counters or inactive material.
Tight photon-detection requirements, 
the ``photon veto" cuts, 
were imposed on the detector subsystems; 
hits in coincidence with the track time within a few nsec and 
with the energy above a threshold of typically 1 MeV were identified 
as the activity due to an extra photon.
The photon veto cuts also ensured that 
only those events 
in which the total photon energy from the $K_{\mu 3\gamma}$ decay
was deposited in the BL 
(and a part of the showers was not recorded in any other subsystem)
were accepted.

\begin{figure}
 \includegraphics[scale=0.45]{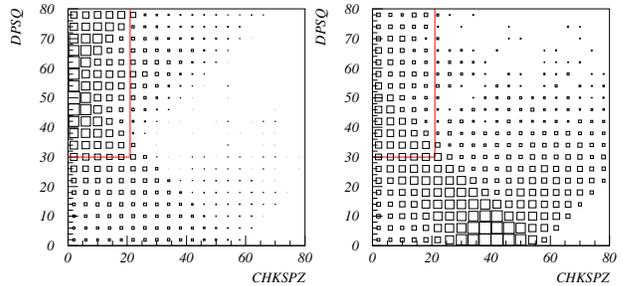}
 \caption{Plots of the variables $DPSQ$ vs. $CHKSPZ$
                for the $K_{\mu3\gamma}$ events (left) 
               and the $K_{\pi 3}$ background events (right)
               generated by the Monte Carlo simulation.}
 \label{fig_CHKSPZ}
 \end{figure}

An extra photon could also go undetected
when two electromagnetic showers in the BL overlapped each other and 
were reconstructed as a single cluster 
(``fused cluster"). 
In the Monte Carlo studies of $K_{\pi 3}$,
it was confirmed that 
the case with the overlap of two photons originated from the same $\pi^0$ was negligible; 
thus the other case, with the overlap of two photons from different $\pi^0$'s
(``odd combination"),
should be considered. 
  We designed two variables
  $DPSQ$ and $CHKSPZ$, described below,
  in which the fused cluster from $K_{\pi 3}$ was characterized. 
Suppose that 
one of the three reconstructed photons 
$\gamma_1$, $\gamma_2$ and $\gamma_3$,
e.g. $\gamma_1$, 
was the fused cluster and
consisted of 
two photons $\gamma_A$ and $\gamma_B$
with the same azimuthal and polar angles
and 
with the energies 
$E_{\gamma_{A}}=\epsilon$ and $E_{\gamma_{B}}=E_{\gamma_{1}}-\epsilon$, 
where $\epsilon$ was a parameter that varied from 0 to $E_{\gamma_1}$.
A pair of invariant masses 
$m_{\gamma_A \gamma_2}$ and $m_{\gamma_B \gamma_3}$
and
$F_{1}\equiv  \sqrt{ ( m_{\gamma_A \gamma_2} - m_{\pi^0} )^2 + 
            ( m_{\gamma_B \gamma_3} - m_{\pi^0} )^2 }$,
            where $m_{\pi^0}$ is the nominal mass of $\pi^0$, 
were calculated as a function of $\epsilon$.
  $F_{1}$ should be small when $\epsilon$ agreed with the correct value.
In the same way, 
$F_{2}$ and $F_{3}$ were calculated 
supposing that $\gamma_2$ and $\gamma_3$ were the fused cluster, 
respectively.
Finally,
 the minimum of $F_{1}$, $F_{2}$ and $F_{3}$
with all the possible values of $\epsilon$ in each hypothesis,
called $DPSQ$ in this analysis, 
was obtained. 
  $DPSQ$ would  be small
  if an  event was from $K_{\pi 3}$
  and one of $\gamma_1$, $\gamma_2$ and $\gamma_3$
  was the fused cluster.
On the other hand, 
the invariant masses of two out of the three reconstructed photons:  
$m_{\gamma_1 \gamma_2}$, 
$m_{\gamma_2 \gamma_3}$ and 
$m_{\gamma_3 \gamma_1}$
were calculated, 
and the minimum of 
$|m_{\gamma_1 \gamma_2}-m_{\pi^0}|$,
$|m_{\gamma_2 \gamma_3}-m_{\pi^0}|$ and
$|m_{\gamma_3 \gamma_1}-m_{\pi^0}|$,
called $CHKSPZ$ in this analysis, 
was obtained.
  If an  event was from $K_{\pi 3}$ with odd combination,  
  the invariant mass of other two photons
  should not be close to $m_{\pi^0}$
  and thereby $CHKSPZ$ would be large.
Figure~\ref{fig_CHKSPZ}
shows the $DPSQ$ vs. $CHKSPZ$ plots
for the $K_{\mu3\gamma}$ events
and 
the $K_{\pi 3}$ background events
(after imposing the BL cuts and the photon veto cuts)
generated by the Monte Carlo simulation.
To suppress the background, 
$DPSQ$ was required to be more than 30 MeV/$c^2$ and 
$CHKSPZ$ was required to be less than 21 MeV/$c^2$.

After imposing the cuts in this subsection, 
4553 events survived.

\subsection{Kinematic fit}
\label{subsec-kinematic}

  A least-squares fit was used
  to improve the resolution of the measured $K_{\mu 3\gamma}$ quantities.
  We assumed just one undetected particle (neutrino) 
  in the final state.
  The kinematic fit was applied to  
  thirteen observables with three constraints.
  The observables were 
     the kinetic energy of the charged particle, 
     the momentum vector of the charged particle, 
     and the momentum vectors of three photons.
  The directions of momentum vectors were defined to be 
    from the kaon decay vertex position.   
  The constraints were: 
  \begin{itemize}
     \item The invariant mass of the total energy and momentum
                should be equal to the nominal mass of $K^+$,
    \item The kinetic energy and momentum of the charged particle should be consistent 
             with a muon hypothesis, and
    \item The invariant mass of one pair of photons should be equal to $m_{\pi^0}$.
  \end{itemize}  
  There are three possible pairings of the three photons to form the $\pi^0$; 
  the combination which minimized the $\chi^2$ of the global kinematic fit
  was chosen among the possible event topologies.
  In simulated $K_{\mu 3\gamma}$ decays, 
  {86\%} of the events that  survived all the selection criteria 
  were reconstructed with the correct pairing.

\begin{figure}
 \includegraphics[scale=0.45]{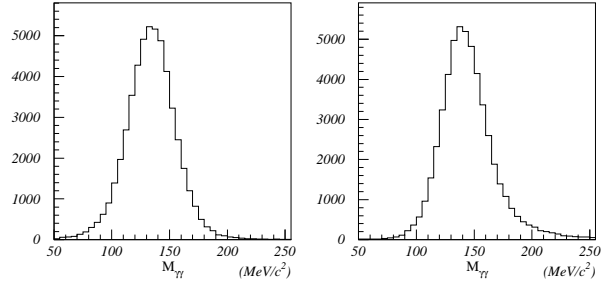}
 \caption{$M_{\gamma\gamma}$ distributions
          of the $\pi^0$ 
          from the $K_{\mu 3}$(left) and $K_{\pi 3}$(right) events.}
 \label{fig_massgg}
 \end{figure}

  The kinematic fit was also applied
  with the assumption of $K_{\mu 3}$ or $K_{\pi 3}$ decay.
  To identify the $K_{\mu 3}$ decay, 
  the best combination of the two photons from the $\pi^0$
  was chosen
  as in the case of the fit to the $K_{\mu 3\gamma}$ assumption, 
  and the remaining photon was ignored. 
  Figure~\ref{fig_massgg}(left) shows
  the invariant $\gamma\gamma$ mass ($M_{\gamma\gamma}$) distribution
  of the $\pi^0$ from the $K_{\mu 3}$ events.
  To identify the $K_{\pi 3}$ decay,
  the events in which four photons were observed in the BL calorimeter were used 
  but the photon with the lowest energy was ignored; 
  the best combination of two photons to form the $\pi^0$
  was chosen  so as to minimize the $\chi^2$.
  Figure~\ref{fig_massgg}(right) shows
  the $M_{\gamma\gamma}$ distribution
  of the $\pi^0$ from the $K_{\pi 3}$ events.
  The results were used to select the events for calibrations
  and to check the performance of Monte Carlo simulation.
  The resolutions used for the muon and the photons
  in  the fit on the $K_{\mu 3\gamma}$  assumption,
  summarized in Table~\ref{kinfit_muon} and Table~\ref{kinfit_gamma},
  were obtained from the studies 
  of the $K_{\mu 3}$ events and the $K_{\pi 3}$ events, respectively. 

  \begin{table}  
  \begin{tabular}{lll}                 \hline \hline
   quantity                                          &   unit          &  resolution \\ \hline
    kinetic energy  $T_{\mu}$         & MeV           & $0.379\times \sqrt{T_{\mu}}$ \\
    momentum  $P_{\mu}$              & MeV/$c$   & $\sqrt{ (0.0227 P_{\mu})^2 - (0.00784 P_{\mu})^4  }$ \\ 
    azimuthal angle $\phi_{\mu}$  & mrad          & $17.6$ \\
    polar angle  $\theta_{\mu}$      & mrad          & $ 32.2\times \cos{\theta_{\mu}}$ \\ 
\hline \hline
 \end{tabular}
     \caption{Resolutions on the muon observables 
      assumed in the kinematic fit to the $K_{\mu 3\gamma}$ hypothesis.}
     \label{kinfit_muon}
   \end{table}
\begin{table}  
  \begin{tabular}{lll}                 \hline \hline
   quantity                                                  &   unit          &  resolution \\ \hline
    energy  $E_{\gamma}$                     & MeV           & $1.61\times \sqrt{E_{\gamma}}$ \\
    azimuthal angle $\phi_{\gamma}$  & mrad          & $39.8$ \\
    polar angle  $\theta_{\gamma}$      & mrad          & $ 65.6\times \cos{\theta_{\gamma}}$ \\ 
\hline \hline
 \end{tabular}
     \caption{Resolutions on the photon observables
      assumed in the kinematic fit to the $K_{\mu 3\gamma}$ hypothesis.}
     \label{kinfit_gamma}
   \end{table}

The $\chi^2$ probability of the kinematic fit on the $K_{\mu 3\gamma}$ assumption, 
$Prob(\chi^2_{K_{\mu3\gamma}})$, was required to be more than  {10\%}. 
In order to suppress 
the $K_{\mu 3}$-1 background, 
the events whose $\chi^2$ probability of the kinematic fit on the $K_{\mu 3}$ assumption, 
$Prob(\chi^2_{K_{\mu 3}})$, was more than {10\%} were discarded. 
After imposing these $\chi^2$ probability cuts, 
360 events survived.

\subsection{Signal region}
\label{subsec-signal}

The following cuts were applied 
to the fitted observables of the surviving events.
The energy of the undetected particle (neutrino) in the final state, $E_{\nu}$,  
should be larger than $60$ MeV
to suppress the $K_{\pi 2 \gamma}$ background. 
The invariant mass of the muon and neutrino
was calculated and, 
in order to suppress the background events with $\pi^+$ decay in flight,  
the invariant mass was required to be larger than 200 MeV/$c^2$.
The polar angle of the neutrino momentum vector 
with respect to the beam axis 
should be  larger than $26^{\circ}$, 
to prevent a photon in the final state
from escaping along the beam direction.
After imposing these cuts, 
178 events survived.

\begin{figure*}
\includegraphics[scale=0.60]{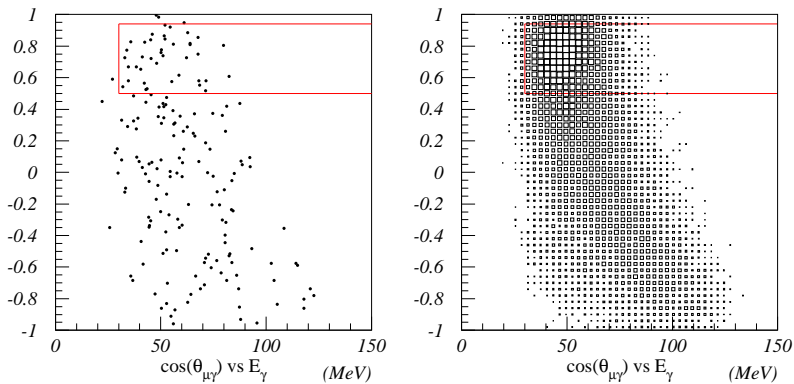}
 \caption{Cosine of $\theta_{\mu\gamma}$ versus $E_{\gamma}$ plots
of the events with all analysis cuts imposed
in real data (left)  and in the sample generated 
by Monte Carlo $K_{\mu3\gamma}$ simulation (right).
The box indicates the signal region.}
 \label{fig_finalscat}
 \end{figure*}

\begin{figure}
 \includegraphics[scale=0.45]{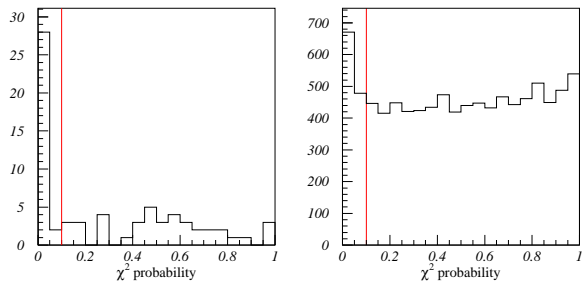}
 \caption{Distributions of  
          $Prob(\chi^2_{K_{\mu3\gamma}})$
          from the events in the signal region
          in real data (left) and in the sample generated 
          by Monte Carlo $K_{\mu3\gamma}$ simulation (right).
          All selection cuts 
          except for the cut $Prob(\chi^2_{K_{\mu3\gamma}})>0.1$
          in Subsection~\ref{subsec-kinematic} are imposed.}
 \label{fig_finalchisq}
\end{figure}

\begin{figure}
 \includegraphics[scale=0.45]{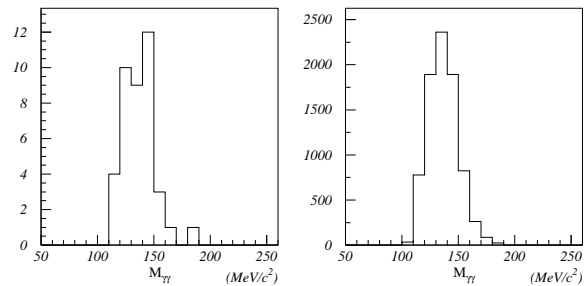}
 \caption{$M_{\gamma\gamma}$ distributions
          of the $\pi^0$ 
          from the events in the signal region 
          in real data (left) and 
          in the sample generated by 
          Monte Carlo $K_{\mu3\gamma}$ simulation (right).}
 \label{fig_finalgg}
\end{figure}

Figure~\ref{fig_finalscat}
shows the cosine of  $\theta_{\mu\gamma}$ versus $E_{\gamma}$ plots
of the events that survived all analysis cuts
in real data and in Monte Carlo $K_{\mu3\gamma}$ simulation.
The signal region was specified with 
$E_{\gamma}>30$ MeV and  $20^{\circ} < \theta_{\mu\gamma} < 60^{\circ}$
  ($0.50 < \cos(\theta_{\mu\gamma}) < 0.94$).
The cut $\theta_{\mu\gamma} < 60^{\circ}$
was required 
to improve the proportion of the correct pairing to $\pi^0$
and to suppress the $K_{\pi 3}$ background.
Figure~\ref{fig_finalchisq} 
and Figure~\ref{fig_finalgg} 
show
the $Prob(\chi^2_{K_{\mu3\gamma}})$ distributions
and the $M_{\gamma\gamma}$ distribution of the $\pi^0$,
respectively, 
from the events in the signal region 
in real data and in Monte Carlo $K_{\mu3\gamma}$ simulation.
Forty events were observed in the signal region.

\section{Background Expectation}
\label{sec-background}

\begin{figure}
\includegraphics[scale=0.45]{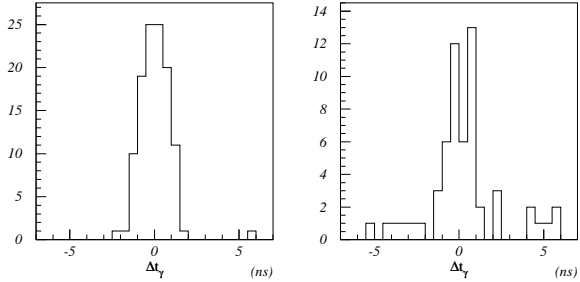}
 \caption{$\Delta {t_{\gamma}}$ distributions
          of the two photons to form the $\pi^0$ (left)
          and of the radiative photon (right) in real data.}
 \label{fig_deltat}
\end{figure}
\begin{figure*}
\includegraphics[scale=0.60]{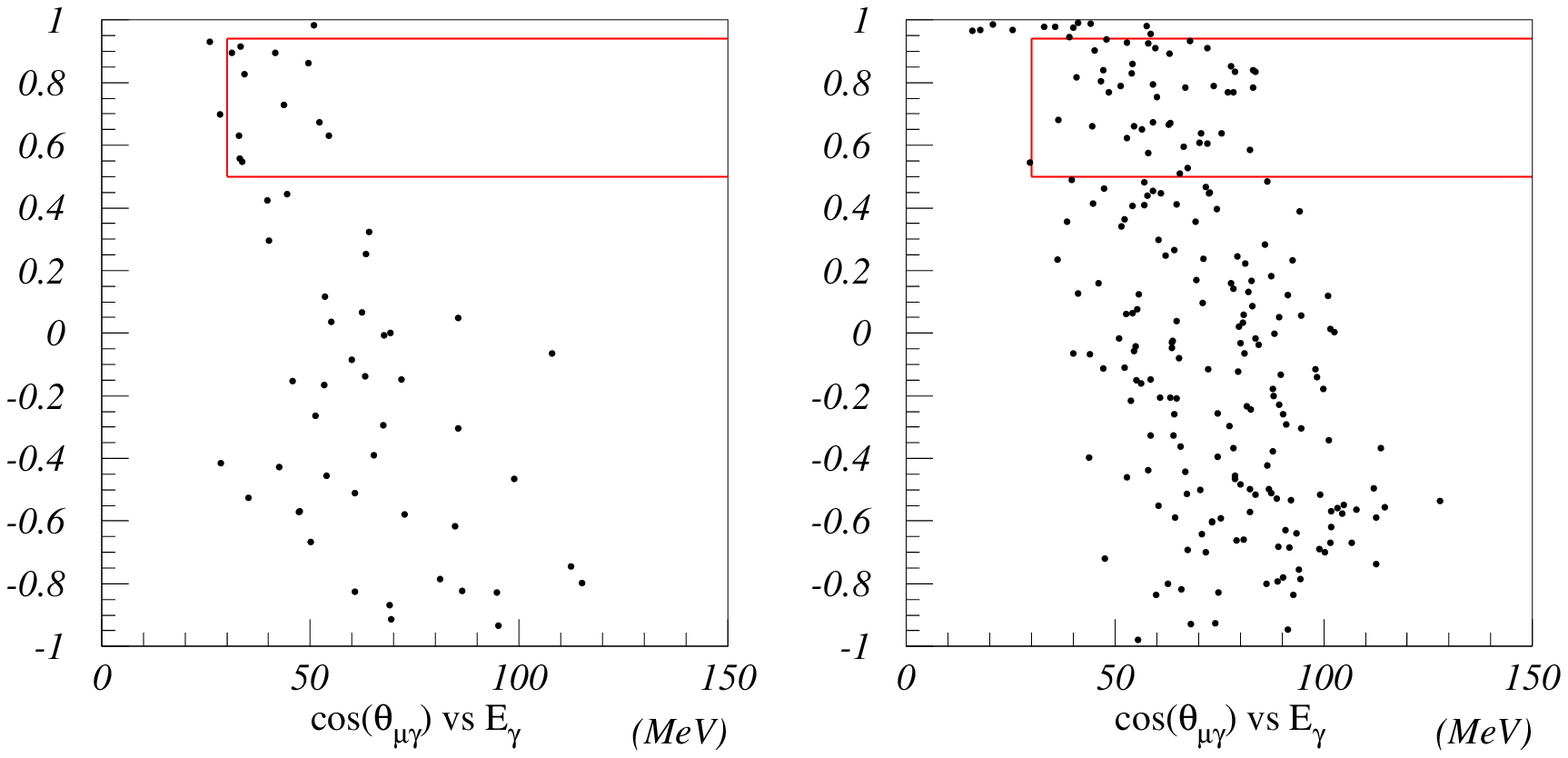}
 \caption{Distributions in 
$\cos (\theta_{\mu\gamma})$ versus $E_{\gamma}$ 
of events selected as 
the $K_{\mu3}$-1(left) and $K_{\pi3}$(right)  background events
as described in the text. 
The box indicates the signal region.}
 \label{fig_bgdscat}
 \end{figure*}

 The $K_{\mu 3}$-1 background 
 was studied
 with real data from the 3$\gamma$ trigger.
 We used the timing of the photons relative to
 the muon track time, $\Delta {t_{\gamma}}$.
 The  $\Delta {t_{\gamma}}$ distributions
 of the two photons to form the $\pi^0$ 
 and of the radiative photon, 
 after imposing  all the offline cuts 
 except for the timing cut to the photons
 ($| \Delta {t_{\gamma}} | <2$ nsec in Subsection~\ref{subsec-selection}),
 are shown in  Figure~\ref{fig_deltat}.
 Figure~\ref{fig_bgdscat}(left) 
 shows the $\cos (\theta_{\mu\gamma})$ versus $E_{\gamma}$ plot
 of the events 
 with a photon 
 in the interval $3$ nsec $< | \Delta {t_{\gamma}} | <6$ nsec 
 in real data; 
 the number of events in the signal region in Fig.~\ref{fig_bgdscat}(left)
 was 11. 
 Assuming that the timing distribution 
 of the BL cluster due to accidental hits was constant,
 the background level of $K_{\mu 3}$-1
 was estimated to be 
 $11 /  (6$nsec$ / 4$nsec$) =
 7.3 \pm 2.2$ events.
%

 Since it was difficult to isolate samples of
 the $K_{\mu 3}$-2 and $K_{\pi 2\gamma}$ backgrounds from the real data, 
 these backgrounds were studied with the samples generated by Monte Carlo simulation. 
 The background level of  $K_{\mu 3}$-2  was estimated to be 
 $< 0.35$ events at the 90\% C.L. and 
 was omitted from the background estimate.
 The background level of $K_{\pi 2\gamma}$ was estimated to be 
 $0.38 \pm 0.08$ events.

The $K_{e 3(\gamma)}$ background was studied 
with the real data.
The $dE/dx$ cuts were inverted 
to enhance the $e^+$ track events, but
no event was left after imposing all the other analysis cuts. 
After taking into account 
the rejection of the $dE/dx$ cuts on $e^+$ tracks
as well as the tagging efficiency of the inverted  $dE/dx$ cuts, 
the background level was estimated to be 
$< 1.1$ events
at the 90\% C.L. 
The background level of $K_{e 3(\gamma)}$ 
was omitted from the background estimate.

The $K_{\pi 3}$ background was studied 
with the real data, in particular with the muon identification cuts 
and the cuts to detect the extra photon.
The studies of photon disappearance
through the inefficiency of the BL, 
the inefficiency of the other subsystems, 
and the fused cluster in the BL 
were made with the BL cuts, the photon veto cuts, and 
the $DPSQ$ and $CHKSPZ$ cuts, respectively.
 Figure~\ref{fig_bgdscat}(right) 
 shows the $\cos (\theta_{\mu\gamma})$ versus $E_{\gamma}$ plot
 of the events 
 after imposing the inverted BL photon-veto cut 
  (detection of the activity due to an extra photon in the BL calorimeter)
 in real data.
 The number of events in the signal region in Fig.~\ref{fig_bgdscat}(right) 
 was 45 and,
 after taking into account 
 the rejection of the offline BL photon-veto cut
 as well as the tagging efficiency of the inverted cut, 
 the background level due to the inefficiency of the BL 
 was estimated to be $7.5 \pm 1.5$ events.
 With the similar method, 
 the background levels
 due to the inefficiency of the other subsystems
 and 
 due to the fused cluster in the BL 
 were estimated to be $0.8 \pm 0.6$ events and
 $0.5 \pm 0.4$ events, respectively.
 The background level of $K_{\pi 3}$ was
 $8.8 \pm 1.6$ events in total.

The background levels
are summarized in Table~\ref{tb_bgdlevel}.
In total, $16.5\pm 2.7$ background events were expected in the signal region.
 \begin{table}  
   \begin{tabular}{lll}                 \hline \hline
   Source                                              &  Background level\ \ \ \ \    &  Samples\\ \hline
   $K_{\mu3}$-1\ \ \ \ \     &   $7.3 \pm 2.2$                   & Data      \\ 
   $K_{\mu3}$-2                 &   $< 0.35$  (90\% C.L.)        & MC       \\ 
   $K_{\pi2\gamma}$                          &  $0.38 \pm 0.08$               & MC       \\ 
   $K_{e3(\gamma)}$                         &  $< 1.1$  (90\% C.L.)           & Data      \\ 
   $K_{\pi3}$                                         &  $8.8 \pm 1.6$ & Data  \\  \hline
  Total                 &   16.5$\pm$2.7 &         \\  \hline \hline
 \end{tabular}
     \caption{Expected background levels in the signal region.
     ``Data" and ``MC" in the rightmost column indicate
     whether real data or Monte Carlo simulation
     were used for the estimation, respectively.   }
     \label{tb_bgdlevel}
   \end{table}

\section{Sensitivity}
\label{sec-sensitivity}

  The acceptance factors for the selection criteria in this measurement
  (Table~\ref{tb_acceptance}) were estimated 
  primarily from Monte Carlo simulation.
  We generated the  $K_{\mu3\gamma}$ sample
  at $O(p^4)$ in ChPT
  with $E_{\gamma}>20$ MeV, 
  which was set to be lower than the offline criteria  on the photon energy~\footnote{
   Thus, the acceptance factors and the single event sensitivity in this section are
   for the $K_{\mu 3\gamma}$ decay  
   in the kinematic region $E_{\gamma}>20$ MeV.
   In Section~\ref{sec-conclusion}, 
   we will first obtain the partial branching ratio
   for $K_{\mu 3\gamma}$ 
   in this kinematic region and then 
   convert it into 
   $BR(K_{\mu3\gamma},\ E_{\gamma} > 30$ MeV and $\theta_{\mu\gamma}>20^{\circ})$
   and 
   $BR(K_{\mu3\gamma},\ 30 < E_{\gamma} < 60$ MeV$)$.}.
  In the simulated events,
  the kinetic energy and momentum of the muon and 
  the energy and $z$ position of the photons were smeared 
  with deviates drawn from a Gaussian distribution 
  to obtain measured resolutions on the observables 
    (Table~\ref{kinfit_muon} and Table~\ref{kinfit_gamma}). 
  The acceptance of the primary selection 
  to the \v{C}erenkov counter, proportional chambers, 
  beam counter
  and the target (Subsection~\ref{subsec-primary})
  and 
  the acceptance loss by the accidental hits in the detector subsystems, 
  in particular the loss due to the veto conditions in the trigger
  (Section~\ref{sec-trigger}),
  were measured with data samples of $K^+\to\mu^+\nu_{\mu}$ decays and
  scattered beam pions, which were
  simultaneously accumulated by the calibration triggers.
  The total acceptance was $\totacctxt$
  and was dominated by the trigger acceptance.
 \begin{table}  
   \begin{tabular}{lll}                 \hline \hline
   Acceptance factors\ \ \ \ \ \ \ \ \ \ \ \ \ \ \ \ \ \ \ \ &        &  Samples\\ \hline
   Muon trigger component                                                &    & \\ 
   \ \ \ \ \ (Section~\ref{sec-trigger})                                         & 0.140                                   & MC  \\
   Photon trigger component                                             &    & \\
   \ \ \ \ \ (Section~\ref{sec-trigger})                                         & 0.00609                               & MC  \\
   Primary selection                                                            &     &  \\
   \ \ \ \ \ (Subsection~\ref{subsec-primary})                          & 0.511                                   & MC, $K_{\mu 2}$\\
   Selection of muon and photons                                    &     & \\ 
   \ \ \ \ \ (Subsection~\ref{subsec-selection})                       & 0.220                                   & MC \\
   Kinematic fit                                                                      &    &  \\
   \ \ \ \ \ (Subsection~\ref{subsec-kinematic})                      & 0.275                                   & MC \\
   Cuts after the kinematic fit                                              &     &  \\
   \ \ \ \ \ (Subsection~\ref{subsec-signal})                             & 0.649                                   & MC \\
   Signal region                                                                    &    &  \\
   \ \ \ \ \ (Subsection~\ref{subsec-signal})                             & 0.351                                   & MC \\ 
   Accidental loss                                                                 &     &  \\
   \ \ \ \ \ (Section~\ref{sec-trigger} and Subsection~\ref{subsec-selection}) 
                                                                      & 0.669                                   & $K_{\mu 2}$, $\pi_{scat}$ \\ \hline
   Total acceptance                            &$\totacc$       &         \\  \hline \hline
 \end{tabular}
     \caption{Acceptance factors for the $K_{\mu3\gamma}$ decay
     in the kinematic region $E_{\gamma}>20$ MeV, and the samples used to determine them.
     ``MC"  in the rightmost column means the sample generated by Monte Carlo simulation.
     ``$K_{\mu 2}$" and ``$\pi_{scat}$" mean 
     the data samples of $K^+\to\mu^+\nu_{\mu}$ decays and scattered beam pions, 
     respectively.}
     \label{tb_acceptance}
   \end{table}

 The single event sensitivity ($SES$) for $K_{\mu3\gamma}$ 
 was derived from the total acceptance, 
 the total exposure of kaons entering the target 
 ($3.5\times 10^{11}$ in Section~\ref{sec-trigger}), 
 and the fraction of kaons entering the target 
 that decayed at rest 
 (called the $K^+$ stop efficiency, $F_{s}$).
 $F_{s}$ was measured to be $0.669\pm 0.014$
 with the events of the IB component of the $K_{\pi 2\gamma}$ decay, 
 which were collected with the 3$\gamma$ trigger   
 and selected~\footnote{
   The triggered events were further prescaled by eight, 
   and were reconstructed 
   with the assumption that the charged track was a $\pi^+$.
   After imposing primary cuts, 
   $\pi^+$ identification cuts, photon selection cuts and 
   photon veto cuts, 
   the kinematic fit with the assumption of the $K_{\pi 2\gamma}$ decay
   was applied. 
   The missing energy should be smaller than $141$ MeV, and  
   the $\pi^+$ momentum $P_{\pi^+}$
   should satisfy
   $140 < P_{\pi^+} < 180$ MeV/$c$.
   A total of 2425 events survived,
   with the background expectation of $3.6$\%. 
   The contribution from the direct photon-emission component  
   of the $K_{\pi 2\gamma}$ decay was estimated to be 2.2\%.}
 with the method
 used in the $K_{\pi 2\gamma}$ analysis of E787
 \cite{E787-kp2g,Tsunemi}. 
 Thus, the sensitivity for $K_{\mu3\gamma}$ was 
 normalized to $K_{\pi 2\gamma}$~\footnote{
   The theoretical prediction for the IB component 
   of the $K_{\pi 2\gamma}$ branching ratio
   in the $\pi^+$ kinetic energy region of 55 to 90 MeV, 
   $2.61\times 10^{-4}$ \cite{DMS-DAFNE},
   was used in this analysis. 
   The Particle Data Group average, 
   $(2.75 \pm 0.15) \times 10^{-4}$ \cite{PDG08}, confirmed this value.} 
 and many systematic uncertainties in the acceptance factors canceled. 
 $F_{s}$ was also measured with 
 the $K_{\pi 3}$ events  collected with the 3$\gamma$ trigger,
 and the results were consistent. 
 We obtained $SES=\sescomb$.

  \begin{table}
  \begin{tabular}{ll}                     \hline \hline
     Estimated systematic uncertainties \ \ \ \ \            &   (\%)   \\ \hline
     BL visible fraction                                                    &    $2.7$  \\
     BL $z$ measurement                                             &     $1.0$  \\
     $e^+$ bremsstrahlung cut                                    &     $1.0$   \\
     Resolution in $T_{\mu}$ and $P_{\mu}$              &    $ 2.8$   \\ 
     Resolution in $E_{\gamma}$                                  &    $ 1.5$   \\ 
     Resolution in photon $z$ position                         &    $1.7$   \\
     Monte Carlo statistical  uncertainty                      &    $1.1$   \\   \hline
     Total systematic uncertainty             &    $\sysuncp$ \\ \hline \hline
   \end{tabular}
   \caption{Systematic uncertainties in the sensitivity and their estimates.}
   \label{tb_sensitivity}
   \end{table}

The systematic uncertainties in $SES$ were due to:
\begin{itemize}
\item The BL calibration of the visible fraction and $z$ measurement, 
  to which the photon selection (Subsection~\ref{subsec-selection}) was sensitive,
\item The difference, between real data and Monte Carlo simulation,
         in the distributions of the variables to select the muon and three photons, 
         in particular the variable on the 
         $e^+$ bremsstrahlung cut (Subsection~\ref{subsec-selection}), and 
\item The smearing of the observables in the Monte Carlo events when the kinematic fit was applied.
\end{itemize}
The systematic uncertainties were studied from the observed sensitivity variation
as each corresponding parameter was varied
over the range that had not been excluded in the calibration.
The statistical uncertainty of the Monte Carlo simulation 
in estimating the acceptance
was also included. 
The systematic uncertainties are 
summarized in Table~\ref{tb_sensitivity}.
By adding them in quadrature, 
the total systematic uncertainty was estimated to be $\pm \sysuncp\%$.

\section{Results and Conclusions}
\label{sec-conclusion}

\begin{figure*}
\includegraphics[scale=0.8]{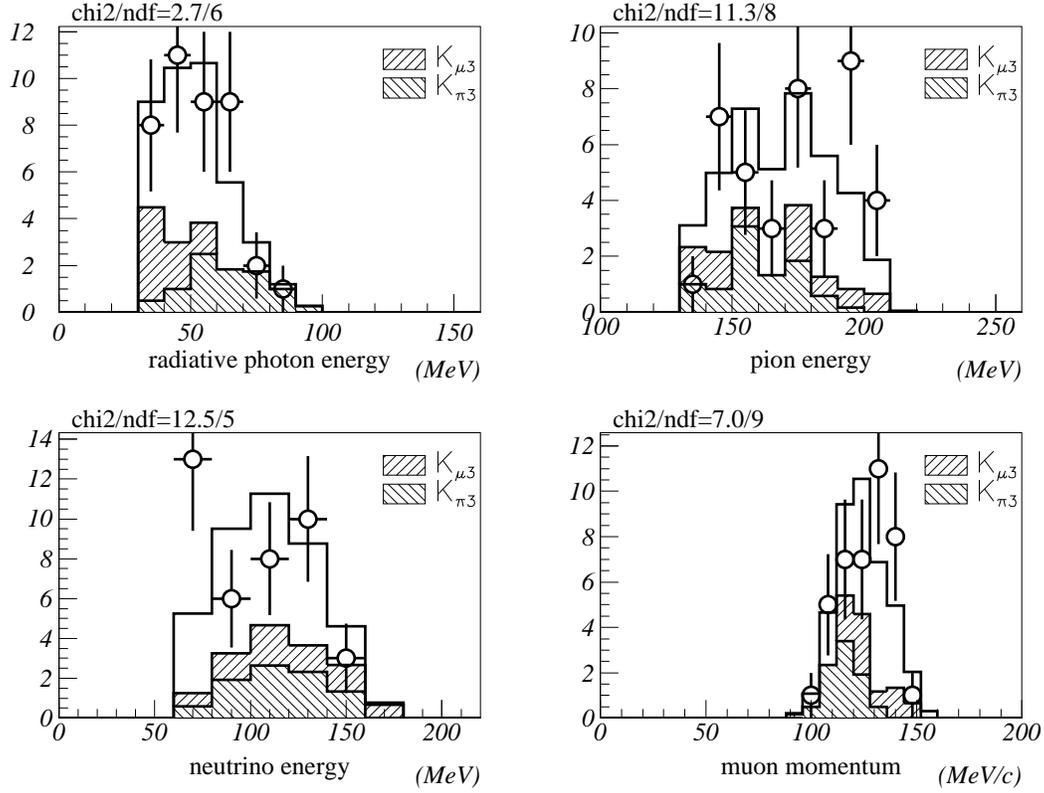}
 \caption{Spectra of the $K^+\to\pi^0\mu^+\nu_{\mu}\gamma$ candidates:
  $E_{\gamma}$ (top left),
 $\pi^0$ energy (top right), 
 $E_{\nu}$ (bottom left) and
 muon momentum (bottom right).
 In each plot, 
 the circles with error bars represent
 the distributions of the 40 events observed in the signal region. 
 The conditions $E_{\gamma}>30$ MeV and $E_{\nu}>60$ MeV have been imposed
 in the analysis.
 The right-hatched and left-hatched histograms represent
 the expected distributions of 
 the $K_{\mu3}$ and $K_{\pi3}$ backgrounds,
 respectively. 
 The unhatched histogram represents 
 the distribution of the Monte Carlo $K_{\mu3\gamma}$ events 
 with the central value of the measured branching ratio
 plus all the background contributions, 
 and should be compared to the circles with error bars.}
 \label{fig_comparison}
 \end{figure*}

From the 40 observed events in the signal region, 
after subtracting the expected background ($16.5\pm 2.7$)
we obtained $23.5\pm 6.9(stat.)$ $K_{\mu 3\gamma}$ events
with the statistical uncertainty of 29\%. 
With $SES=\sessys$, 
the partial branching ratio for $K_{\mu 3\gamma}$ 
in the kinematic region $E_{\gamma}>20$ MeV was 
$\brses$.
The kinematic distributions of the observed 40 events 
were compared to the spectra  
predicted from the $K_{\mu3\gamma}$ decay plus all the background contributions, 
as shown in  Fig.~\ref{fig_comparison}.
The $\chi^2/$degree-of-freedom values 
to evaluate  the match between data and predictions were 
$2.7/6$, $11.3/8$, $12.5/5$ and $7.0/9$
in the distributions of $E_{\gamma}$, $\pi^0$ energy, $E_{\nu}$ and 
muon momentum, respectively.

The factors $0.628$ and $0.437$,  
estimated from the theoretical $K_{\mu3\gamma}$ spectrum  
at $O(p^4)$ in ChPT,
are used to convert 
the measured partial branching fraction
into
$BR(K_{\mu3\gamma},\ E_{\gamma} > 30$ MeV and $\theta_{\mu\gamma}>20^{\circ})$
and 
$BR(K_{\mu3\gamma},\ 30 < E_{\gamma} < 60$ MeV$)$, 
respectively.
Finally,  the results of the measurement were: 
\begin{center}
\begin{tabular}{l}
 $BR(K_{\mu3\gamma},\ E_{\gamma} > 30$ MeV and $\theta_{\mu\gamma}>20^{\circ})$ \\
 $=\ \brfinal$ \\
\end{tabular}
\end{center}
and
\begin{center}
\begin{tabular}{l}
\\
$BR(K_{\mu3\gamma},\ 30 < E_{\gamma} < 60$ MeV$)$ \\
 $=\ \brpdg$. \\
 \\
\end{tabular}
\end{center}
The $BR(K_{\mu3\gamma},\ E_{\gamma} > 30$ MeV and $\theta_{\mu\gamma}>20^{\circ})$
was consistent with the theoretical prediction 
$2.0\times 10^{-5}$ \cite{BEG-DAFNE, BEG93}
as well as 
with the previous measurement
$(2.4\pm 0.5(stat.) \pm 0.6(syst.)) \times 10^{-5}$ in \cite{Shimizu06}, 
and the $BR(K_{\mu3\gamma},\ 30 < E_{\gamma} < 60$ MeV$)$
was consistent with the previous measurement 
 $(1.5\pm 0.4) \times 10^{-5}$ in  \cite{PDG08}.
The results of the E787 measurement provide better precision than
the previous results.


\section*{Acknowledgments}

We acknowledge the contributions made by colleagues 
who participated in earlier phases of the E787 experiment, 
including 
M.~Atiya, T.F.~Kycia(deceased), D.~Marlow, and A.J.S.~Smith. 
We gratefully acknowledge the dedicated effort of the technical staff 
supporting this experiment and of the Brookhaven Collider-Accelerator Department. 
This research was supported in part by the U.S. Department of Energy 
under Contracts No. DE-AC02-98CH10886, No. W-7405-ENG-36,
and Grant No. DE-FG02-91ER40671, by the Ministry of Education, 
Culture, Sports, Science and Technology
of Japan through the Japan-US Cooperative Research 
Program in High Energy Physics and under the Grant-in-Aids for Scientific 
Research, for Encouragement of Young Scientists and for JSPS Fellows, and by 
the Natural  Sciences and Engineering Research Council and the National 
Research Council of Canada.

\end{document}